\documentclass[11pt,a4paper]{article}
\pdfoutput=1
\usepackage{authblk}
\usepackage{amsmath,amssymb,amsfonts}
\usepackage{algorithmic}
\usepackage{graphicx}
\usepackage{textcomp}
\usepackage{xcolor}
\def\BibTeX{{\rm B\kern-.05em{\sc i\kern-.025em b}\kern-.08em
    T\kern-.1667em\lower.7ex\hbox{E}\kern-.125emX}}

\usepackage{lipsum}
\usepackage{subfiles}

\usepackage[utf8]{inputenc}
\usepackage{physics,mathtools}

\usepackage[
    hyperfootnotes=false,
    colorlinks=true,
    linkcolor=blue!50!black,
    urlcolor=blue!50!black,
    citecolor=blue!50!black,
    anchorcolor=blue!50!black,
    pagebackref=false,
    unicode=true]{hyperref}
    
\usepackage[left=2cm,right=2cm,top=2cm,bottom=2cm]{geometry}

\newtheorem{remark}{Remark}

\newcommand{\cS}{\mathbf{S}}
\newcommand{\cL}{\mathbf{L}}
\newcommand{\cP}{\mathbf{P}}
\newcommand{\cI}{\mathbf{I}}
\newcommand{\cA}{\mathbf{A}}
\newcommand{\cH}{\mathbf{H}}
\newcommand{\cR}{\mathbf{R}}
\newcommand{\cD}{\mathbf{D}}
\newcommand{\cV}{\mathbf{V}}

\begin{document}
\author[1]{Balázs Csutak} 
\author[1]{Péter Polcz}
\author[1]{Gábor Szederkényi\footnote{corresponding author}}
\affil[1]{\small Pázmány Péter Catholic University, Faculty of Information Technology and Bionics, Práter u. 50/a,  H-1083 Budapest, Hungary} 
\date{}

\title{Computation of COVID-19 epidemiological data in Hungary using dynamic model inversion
\thanks{P. Polcz gratefully acknowledges the support of the New National Excellence Program scholarship (ÚNKP-20-4-I-PPKE-41).
B. Csutak was supported by the UNKP-20-3-I-PPKE-66 project of the New National Excellence program of the Ministry for Innovation and Technology.
G. Szederkényi acknowledges the partial support by the European Union, co-financed by the European Social Fund through the grant EFOP-3.6.3-VEKOP-16-2017-00002.
This work was partially supported by the National Research, Development, and Innovation Office through the grant NKFIH-OTKA-131545.}
}


\maketitle

\begin{abstract}
In this paper, we estimate epidemiological data of the COVID-19 pandemic in Hungary using only the daily number of hospitalized patients, and applying well-known techniques from systems and control theory. We use a previously published and validated compartmental model for the description of epidemic spread. Exploiting the fact that an important subsystem of the model is linear, first we compute the number of latent infected persons in time. Then an estimate can be given for the number of people in other compartments. From these data, it is possible to track the time dependent reproduction numbers via a recursive least squares estimate. The credibility of the obtained results is discussed using available data from the literature.
\end{abstract}


\section{Introduction}
Since the spring of 2020, the COVID-19 pandemic has put an enormous burden on the societies, economies and healthcare systems of most countries. It is of fundamental importance to track the evolution of the epidemic process in order to prepare healthcare capacities, or to design efficient measures and vaccination policies against the disease spread. It is well-known that only a fraction of the actually infected persons get registered into the official databases. This ratio varies widely between countries depending on the testing activity and related policies. E.g., as of
Apr. 28., 2021, the official data for the cumulative numbers of infected in Czechia and Hungary are 1.626.033 \cite{Cseh_confirmed_cases} and 774.399 \cite{koronavirus.gov.hu_confirmed_cases}, respectively.
Considering the similarities between these two countries in terms of geographic location, area, population size, the stringency of interventions, the organization of healthcare, and the number of reported COVID-deaths, it is highly unlikely that there could be such a huge difference in the true numbers. There exist several attempts in the literature for the reconstruction of real epidemic data. In \cite{phipps2020robust} a robust statistical estimate with uncertainty analysis was given for the detection rate in several countries. It turns out from the published results that during the 2020 spring wave of the pandemics, the detection rate could be as small or even lower than  10\% for e.g., Sweden, Italy, Spain, or may have reached 40-50\% in the case of Australia and South Korea, where huge effort was put into immediate contact tracing and testing. Somewhat tighter upper bounds were computed for the possible total number of cases in several European countries in \cite{rocchetti2020estimating} using a data-driven estimator from the distributions of daily cases and deaths. It is clear from the above that the recorded number of infected people alone is generally not a reliable source for the tracking of the epidemics. Moreover, it is also visible from the Hungarian data that recoveries were not followed and recorded precisely in the second wave of the pandemic until the middle of December, 2020 (see, e.g. at \cite{HungaryC59:online}). Therefore, the epidemic curve showing the active cases cannot reflect the real situation even in a qualitative way. After the spring wave of COVID-19, a representative serological test was designed and carried out in Hungary which estimated that exposure was approximately 0.68\% among the total population by the middle of May, 2020 \cite{merkely2020novel}. However, no similar study has been done since the summer of 2020, although it would certainly have huge significance in exploring the current situation and refining our models.

There is a huge literature on epidemic modeling and recently on the modeling of the COVID-19 pandemic for different purposes such as identification, control, or prediction \cite{shinde2020forecasting}. E.g., in \cite{pinter2020covid}, a hybrid machine learning approach is proposed to predict the number of infected individuals and the mortality rates of the first wave of the COVID-19 outbreak in Hungary. In \cite{ardabili2020covid} the authors show that epidemic outbreak prediction can be efficiently supported by integrating machine learning and SEIR models.

In systems and control theory, it is a common practice to estimate unknown quantities (e.g., states or parameters) from measurement data using dynamical models of the observed phenomena \cite{sontag2013mathematical,lennart1999system}. Population-level deterministic epidemic models are most often written in a nonnegative compartmental form with polynomial nonlinearities representing the infection mechanism and linear subsystems describing the transitions between certain compartments \cite{brauer2012mathematical}. We will use a control oriented ODE model of the pandemics in Hungary which was proposed and used in \cite{2020_Peni.etal}, and was derived from the more detailed model in \cite{rost2020early}.

Based on the above, the purpose of this paper is to propose a model-based computation approach to study the second and third waves of the COVID-19 pandemic in Hungary. Due to the previously mentioned unreliability of other data sources, we only use the number of hospitalized people available from \cite{Dataonho54:online} as input data, assuming that testing is wide-spread and quick enough in hospitals.

\section{The applied compartmental model}
\label{sec:Model}

\subsection{Model description}
\label{sec:Model_desc}
\begin{figure}[h]
    \centering
    \includegraphics[width=0.8\linewidth]{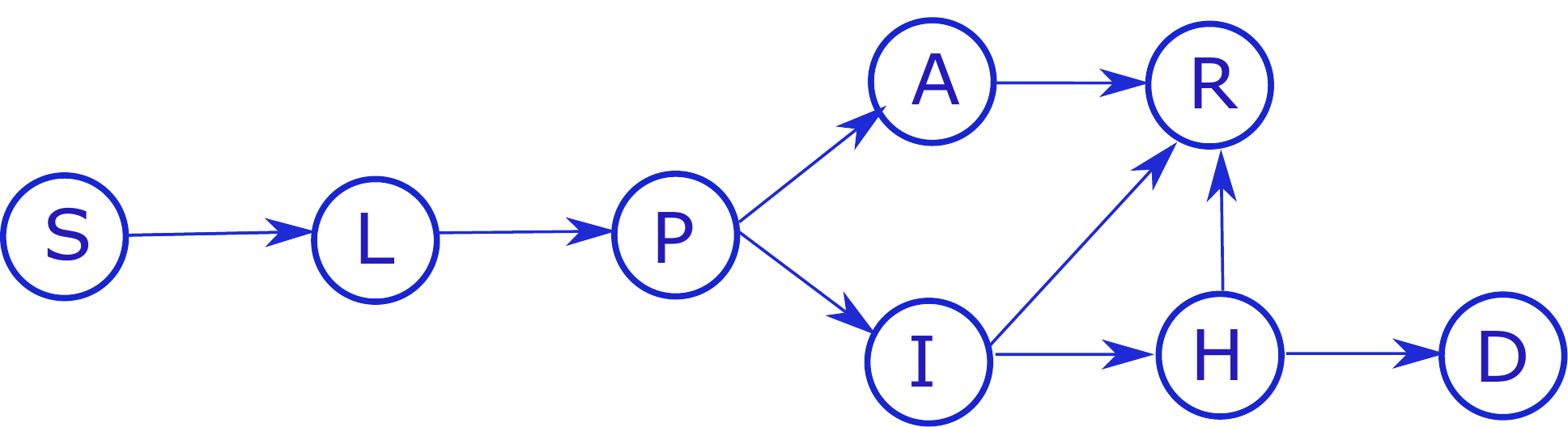}
	\caption{
	    \label{fig:diagram}
	    Transition diagram of the disease spreading model.
	    Circles represent compartments and arrows represent
transitions between these compartments.
	}

\end{figure}

To describe the transmission dynamics of the disease, we use a compartmental model describing the spreading characteristics of COVID-19. We divide the population of $N$ individuals into eight classes, representing the different stages of the illness, discussed in detail in \cite{2020_Peni.etal}.
The compartments used stand for the following: the susceptibles ($\cS$) are those individuals, who can be infected by the disease (i.e., neither have been infected  nor yet obtained immunity by, eg., vaccination). Latents ($\cL$) are infected people in the very first stage, lacking any symptoms and also incapable to transmit the pathogen. This is followed by the pre-symptomatic ($\cP$) stage, including individuals already infectious, but still without symptoms. As a significant part of the infected show no characteristic symptoms, and thus they being infected is often not confirmed, we handle the asymptomatic ($\cA$) and symptomatic infected ($\cI$) classes differently. Members of $\cA$ always recover ($\cR$) after a certain time, but for some part of $\cI$ hospitalization is needed. Hospitalized patients ($\cH$) may recover or decease ($\cD$). The possible transitions between the compartments are illustrated in Fig. \ref{fig:diagram} and their dynamics is described formally by the ODEs below.
\begin{subequations}
    \label{eq:compartmental}
    \begin{align}
\label{eq:compS}
            &\dot \cS(t) = - \beta \left[
                \cP(t) + \cI (t)  +
                \delta \cA(t) \right] \cS(t) / N,
            \\
            \label{eq:compL}
            &\dot \cL(t) =   \beta \left[
                \cP(t) + \cI(t)  +
                \delta \cA(t) \right] \cS(t) / N- \alpha \cL(t),
            \\
            \label{eq:compP}
            &\dot \cP(t) =   \alpha \cL(t) - p \cP(t),
            \\
            \label{eq:compI}
            &\dot \cI(t) =   q p \cP(t) - \rho_I \cI(t),
            \\
            \label{eq:compA}
            &\dot \cA(t) =   (1-q) p \cP(t)  - \rho_A \cA(t) ,
            \\
            \label{eq:compH}
            &\dot \cH(t) =  \rho_I \eta \cI(t) -h \cH(t),
            \\
            \label{eq:compR}
            &\dot \cR(t) =   \rho_I (1-\eta) \cI(t) + \rho_A \cA(t)+ (1-\mu)h \cH(t),
            \\
            \label{eq:compD}
            &\dot \cD(t) = \mu h \cH(t).
\end{align}
\end{subequations}
From now on, we suppress the time arguments in the state variables as it is commonly done in the literature.

\subsection{Model parameters}
\label{sec:Model_params}
We set the model parameters according to the estimates published in related literature.
We used the following model parameters for \eqref{eq:compartmental} which had been determined  by updating the parameter set published in \cite{2020_Peni.etal} with newer Hungary-specific data. The population of Hungary is $N=9.8\cdot 10^6$ people.
The latent period ($\alpha^{-1}$) is 2.5 days on average, whereas, the pre-symptomatic period ($p^{-1}$) is generally 3 days.
The infectious period of both the symptomatic ($\rho_I$) and asymptomatic ($\rho_A$) classes is approximately 4 days.
The average length of hospitalization ($h^{-1}$) before recovery or death is 10 days. The hospitalization probability of the symptomatic ($\eta$) is $0.076$.
The probability of developing symptoms ($q=0.6$) and the relative infectiousness of the asymptomatic ($\delta=0.75$) are based on CDC estimates. The death ratio among hospitalized is $\mu=0.185$.

The basic reproduction number, expressing the average number of new infections generated by a single infected individual in a fully susceptible population, is given as
\begin{equation}\label{R0}
    R_0=\beta \left( \frac{1}{p} +\frac{q}{\rho_I}+\frac{\delta(1-q)}{\rho_A}\right).
\end{equation}
Using $R_0=2.2$ estimated from early Hungarian data, and the already determined parameters, a nominal value of $\beta=1/3$ can be derived, but note that this parameter largely depends on the actual restrictions and conditions, therefore, it will be estimated as a time-varying parameter. The time-dependent reproduction number for our model taking into account the change of $\beta$ and the decrease of the susceptible population is given by
\begin{align}
    \label{eq:Rct}
    R_c(t) = \beta(t) \left( \frac{1}{p} +\frac{q}{\rho_I}+\frac{\delta(1-q)}{\rho_A}\right)\frac{\cS(t)}{N}.
\end{align}

\section{Computation methodology}

\subsection{Data processing}
\label{sec:Comp_preproc}
Dynamic inversion is known to be sensitive to the abrupt changes in the derivatives of measured signals caused by fluctuations and noise. Therefore, we applied the following preprocessing to smoothen the input data.

Let $\check y_k$ denote the number of hospitalized patients on day $k$ collected from \cite{Dataonho54:online}, $k = 1,\dots,T$.
As the starting date ($k = 1$), we considered the 20th of August, 2020.
The available data was collected up till the 28th of April, 2021.

Generally, at weekends the healthcare system has a decreased capacity to perform tests and document the confirmed cases.
Therefore, as it is commonly done in the literature and engineering practice, we apply a $7$-day long moving average filter to the officially published hospitalization data record.
Formally, we consider the following smoothed time series of the numbers of hospitalized patients:
\begin{align}
    \textstyle
    \bar y_k = \frac17 \sum_{i = -3}^{3} \check y_{k+i}, \text{ if } k = 4,\dots,T-3.
\end{align}
Obviously, at the two ends of the time series, the sliding window has to be truncated as follows:
\begin{align}
    \textstyle
\bar y_k = \frac{1}{\min(3,T-k)+\min(3,k)+1} \sum_{i = -\min(3,k)}^{\min(3,T-k)} \check y_{k+i}.
\end{align}

Next, we used a cubic spline interpolation technique, ensuring the result being twice continuously differentiable. We divided the data series into $n=15$ equally long segments, the endpoints  being at $t_i=i\cdot(T/n)$, $t=0,1,\dots,n$ in time. As these time points (expressed in days) are not necessarily integers, we used a linear interpolation to calculate $\overline y _{t_i}$ from $\overline{y}_{\lfloor i\cdot(T/n)\rfloor}$ and $\overline{y}_{\lfloor i\cdot(T/n)\rfloor+1}$. The spline was fitted on data points $(t_i, \overline y _{t_i})$, $i=0,1,\dots,n$
using not-a-knot conditions for the slope endpoints.
Let the spline interpolated hospitalization data be denoted by $y_t$, $t \in [1,T]$.
The original measurement ($\check y$) and the results ($\bar y$ and $y$) of the smoothing process can be seen in Fig. \ref{fig:smoothing}.
Additionally, Fig. \ref{fig:derivaties_of_H} illustrates that the data preprocessing steps make it possible to compute the first three derivatives of $y$ relatively noise-free.
Henceforth, in all computations, we use $y$ as the single available information about the time evolution of the epidemic spread.

\color{black}

\newcommand{\SigmaPIH}{\Sigma_{\mathrm{\cP\cI\cH}}}
\newcommand{\LSA}{M_a}
\newcommand{\LSB}{M_b}

\subsection{Input computation by inversion}
\label{sec:Comp_inv_LS}

Consider the $(\cP,\cI,\cH)$ subsystem of \eqref{eq:compartmental} in the following continuous-time linear time-invariant (LTI) state-space form:
\begin{align}
    \label{eq:PIH_ss}
    \SigmaPIH: &\begin{dcases}
        \dot x = A x + B u, \\
        y = C x,
    \end{dcases}
    ~~\text{ where }
    x = \spmqty{ \cP \\ \cI \\ \cH },~
    u = \cL,
    \\
    \label{eq:PIH_ss_ABC}
    &~~A = \spmqty{
       -p   &  0           &  0 \\
        p q & -\rho_I      &  0 \\
        0   &  \eta \rho_I & -h \\
    },~
    B = \spmqty{
        \alpha  \\
        0       \\
        0       \\
    },~
    C = \qty( 0 ~ 0 ~ 1 ).
\end{align}
The input of $\SigmaPIH$ is the daily number of people in the latent phase ($u = \cL$), whereas, the output is \emph{(the filtered time series of)} the daily number of hospitalized patients ($y = \cH$).
The continuous-time transfer function of $\SigmaPIH$ is the following:
\begin{align}
    \label{eq:PIH_CT_tf}
    G(s) = \frac{0.00152}{s^3 + 0.6833 s^2 + 0.1417 s + 0.008333},
\end{align}
from which we can see that the analytical inverse of the system is non-causal.

To compute the unknown input of $\SigmaPIH$, we consider the discrete-time (DT) model of $\SigmaPIH$ with $T_s = 1 \text{ (day)}$ sampling period.
Using zero order hold, the transfer function of the discretized model is the following:
\begin{align}
    \label{eq:PIH_DT_tf}
    W(z) = \frac{
        b_2 z^2 + b_1 z + b_0
    }
    {
        z^3 + a_2 z^2 + a_1 z + a_0
    },
\end{align}
where
$a_0 = - 0.5049$,
$a_1 = 1.911$,
$a_2 = - 2.4$,
$b_0 = 1.52  \cdot 10^{-4}$,
$b_1 = 7.225 \cdot 10^{-4}$,
$b_2 = 2.139 \cdot 10^{-4}$.
The corresponding DT input-output model can be written as follows:
\begin{align}
    \label{eq:PIH_DT_io}
    y_{k\!+\!3} \!+\! a_2 y_{k\!+\!2} \!+\! a_1 y_{k\!+\!1} \!+\! a_0 y_{k} \!=\! b_2 u_{k\!+\!2} \!+\! b_1 u_{k\!+\!1} \!+\! b_0 u_{k}.
\end{align}
First, we compute the input from the output using a standard least squares deconvolution. Using the available filtered time series of $y_{k = 1,\dots,T}$, the input-output difference equation \eqref{eq:PIH_DT_io} can be written in the following matrix-vector format:
\newcommand{\sdd}{\scalebox{0.5}{$\ddots$}}
\begin{align}
    \label{eq:PIH_DT_io_lineq}
    &\LSA \mathbf{y} = \LSB \mathbf{u}, \text{ where }
\\
    &\LSA \!\!=\!\! \spmqty{
        a_{0} & a_{1} & a_{2} & 1     & 0     & \dots & \dots \\
        0     & a_{0} & a_{1} & a_{2} & 1     & 0     & \dots \\
        \sdd  & \sdd  & \sdd  & \sdd  & \sdd  & \sdd  & \sdd  \\
\dots & \dots & 0     & a_{0} & a_{1} & a_{2} & 1     \\
},
    \LSB \!\!=\!\! \spmqty{
        b_{0} & b_{1} & b_{2} & 0     & 0     & \dots \\
        0     & b_{0} & b_{1} & b_{2} & 0     & 0     \\
        \sdd  & \sdd  & \sdd  & \sdd  & \sdd  & \sdd  \\
\dots & \dots & 0     & b_{0} & b_{1} & b_{2} \\
},
    \\
    &\mathbf{y} = \pqty{ y_1 ~ \dots ~ y_{T} }^\top,~
    \mathbf{u} = \pqty{ u_1 ~ \dots ~ u_{T-1} }^\top.
\end{align}
The system of linear equations in \eqref{eq:PIH_DT_io_lineq} is under-determined in $\mathbf{u}$, since $\LSA \in \mathbb{R}^{{(T - 3)}\times{T}}$ and $\LSB \in \mathbb{R}^{{(T - 3)}\times{(T - 1)}}$.
The least-squares solution for $\mathbf{u}$ can be computed by considering the Moore--Penrose right pseudo-inverse $\LSB^\ddagger = \LSB^\top (\LSB \LSB^\top)^{-1}$ of the row-shaped matrix $\LSB$, namely:
\begin{align}
    \label{eq:PIH_DT_io_lineq_LS_sol}
    \check{\mathbf{u}}_{\mathrm{ls}} = \LSB^\ddagger \LSA \mathbf{y}.
\end{align}

After computation, we filtered the time series $\check{\mathbf{u}}_{\mathrm{ls}}$ with a $7$-day moving average filter, which resulted in $u_{\mathrm{ls},k}$, $k = 1,\dots,T-1$. Having an estimate of the input of the system $\SigmaPIH$, we can apply a standard linear state observer to estimate the population in the non-measured compartments $(\cP,\cI,\cA,\cR)$. An estimate for $\cD$ can be computed from \eqref{eq:compD}, where the official governmental data can be used as well. However, we do not use the published data on COVID deaths for inversion, and the dynamics of the other compartments do not depend on $\cD$.

\subsection{Unknown-input state estimation}
\label{sec:Comp_inv_uio}
As another possible solution, in this subsection we present an unknown input observer approach for system inversion and state estimation.
It can be seen from the transfer function model \eqref{eq:PIH_CT_tf} that
the LTI subsystem $\SigmaPIH$ of \eqref{eq:compartmental} has relative degree $3$, i.e., the input signal does appear only in the third (and higher order) derivatives of the output equation:
\begin{align}
    \label{eq:PIH_d1y}
    &\dot y = C A x + C B u = C A x, && (C B = 0)
    \\
    \label{eq:PIH_d2y}
    &\ddot y = C A^2 x + C A B u = C A^2 x, && (C A B = 0)
    \\
    \label{eq:PIH_d3y}
    &\dddot y = C A^3 x + C A^2 B u.
\end{align}
We can check that $\SigmaPIH$ in itself does not satisfy the necessary criteria of
\cite[Lemma 3.1]{Chen.Patton_1999} or \cite[Criterion 2]{2001_Moreno}
for the existence of an \emph{unknown-input observer} which may seem contradictory knowing the result of the previous subsection.

Observe that, if we consider \eqref{eq:PIH_d1y} and \eqref{eq:PIH_d2y} as two additional outputs of $\SigmaPIH$, then
the augmented state-space model
\begin{align}
    \label{eq:PIH_ss_aug}
    \begin{dcases}
        \dot x = A x + B u,
        \\
        \mathcal Y = \mathcal O_3 \, x,
    \end{dcases}
    ~~\text{ where } \mathcal Y = \spmqty{ y \\ \dot y \\ \ddot y },~ \mathcal O_3 = \spmqty{C \\ C A \\ C A^2}
\end{align}
is unknown-input observable.
Furthermore, \eqref{eq:PIH_d3y} allows us to algebraically express the unknown input from the \emph{third} derivative of signal $y$.

Following the ideas of \cite[Section 3.2.1]{Chen.Patton_1999}, we construct the following unknown-input observer for \eqref{eq:PIH_ss_aug}:
\begin{align}
    \label{eq:PIH_uio_ss}
    \begin{dcases}
        \dot z = F z + K \mathcal Y,
        \\
        x_{\mathrm{uio}} = z + H \mathcal Y,
        \\
        u_{\mathrm{uio}} = (C A^2 B)^{-1}\pqty{ \dddot y - C A^3 x_{\mathrm{uio}} }.
    \end{dcases}
\end{align}
The coefficient matrices of \eqref{eq:PIH_uio_ss} are:
\begin{align}
    \begin{aligned}
        &K = K_1 + K_2,~ H = B (\mathcal O_3 B)^\dagger,
        \\
        &F = A - H \mathcal O_3 A - K_1 \mathcal O_3,~ K_2 = F H,
    \end{aligned}
\end{align}
where
$(\mathcal O_3 B)^\dagger = \pqty\big{(\mathcal O_3 B)^\top (\mathcal O_3 B)}^{-1} (\mathcal O_3 B)^\top$ constitutes the Moore--Penrose left pseudo-inverse of matrix $\mathcal O_3 B$, and
$K_1$ is selected such that $F$ is a stability matrix.

To summarize, the non-measured state variables $(\cP,\cI)$ of $\SigmaPIH$ can be reconstructed using the first two derivatives of $y$, whereas, the input reconstruction requires also the third derivative of $y$.
The remaining state variables $(\cA,\cR,\cD)$ of \eqref{eq:compartmental} can be reconstructed by considering the estimated trajectories of $(\cP,\cI)$ and the measured trajectory of $y$. Finally, using the fact that the whole population is conserved in our compartmental model, the number of susceptibles can be estimated simply as
\begin{align}
\cS=N-(\cL + \cP + \cI + \cA + \cH + \cR + \cD + \cV),
\end{align}
where $\cV$ denotes the cumulative number of people how have become immune due to vaccination.

\begin{remark}
    \label{TMP_Tovu0ltLdm3r}
    In the computation, we considered a simple vaccination model taking into consideration the documented efficacy of vaccines applied in Hungary. Based on this, we assumed that 85\% of vaccinated people do not take part in the infection process any more from 21 days after administering the first dose.
The time series of the number of vaccinated people were collected from the official website \cite{koronavirus.gov.hu} of the Government of Hungary.
\end{remark}

\begin{small}
\newlength{\tilelength}
\setlength{\tilelength}{0.45\textwidth}
\begin{figure*}
    \begin{minipage}{\tilelength}
        \centering
        \includegraphics[width=\textwidth]{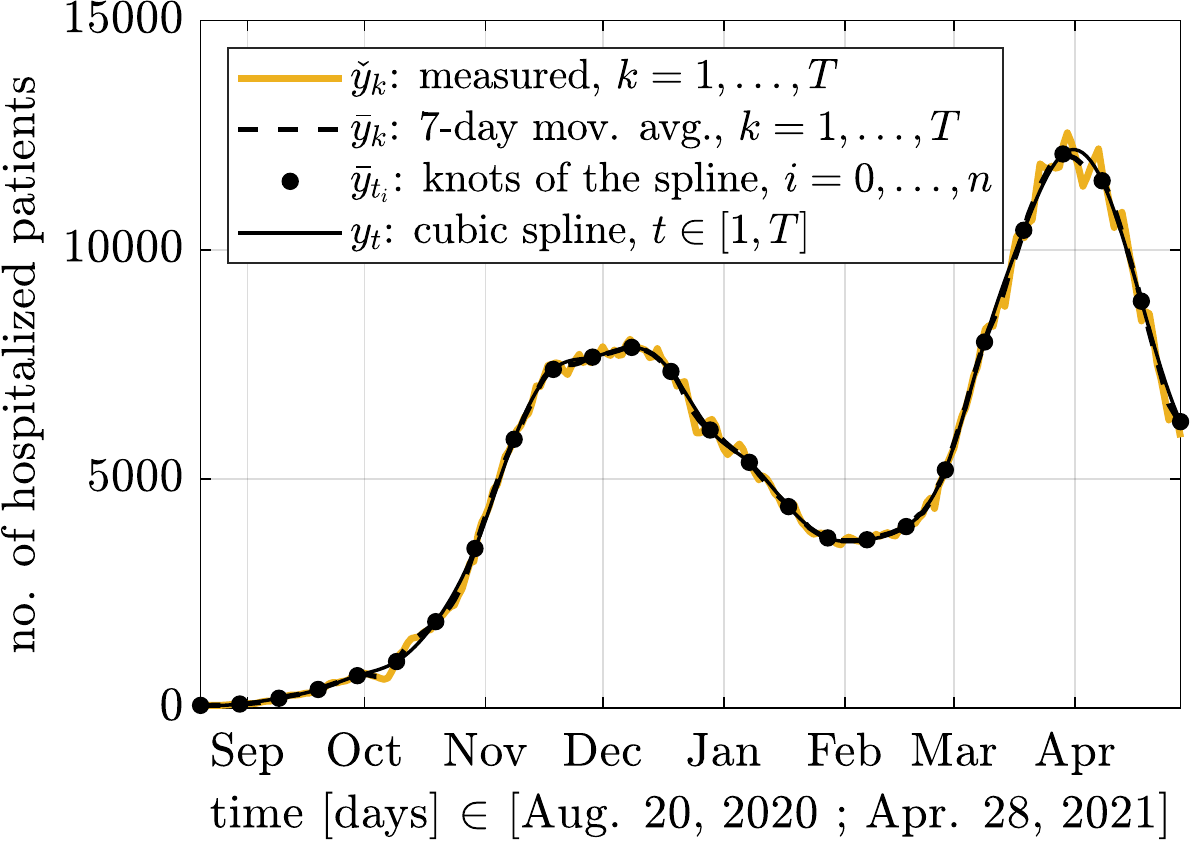}
        \caption{Smoothed time series (solid black) compared to the officially published hospitalization data (thick yellow) and the 7-day long moving average filtered data (dashed black).}
        \label{fig:smoothing}
    \end{minipage}
    \hfill
    \begin{minipage}{\tilelength}
        \centering
        \includegraphics[width=\textwidth]{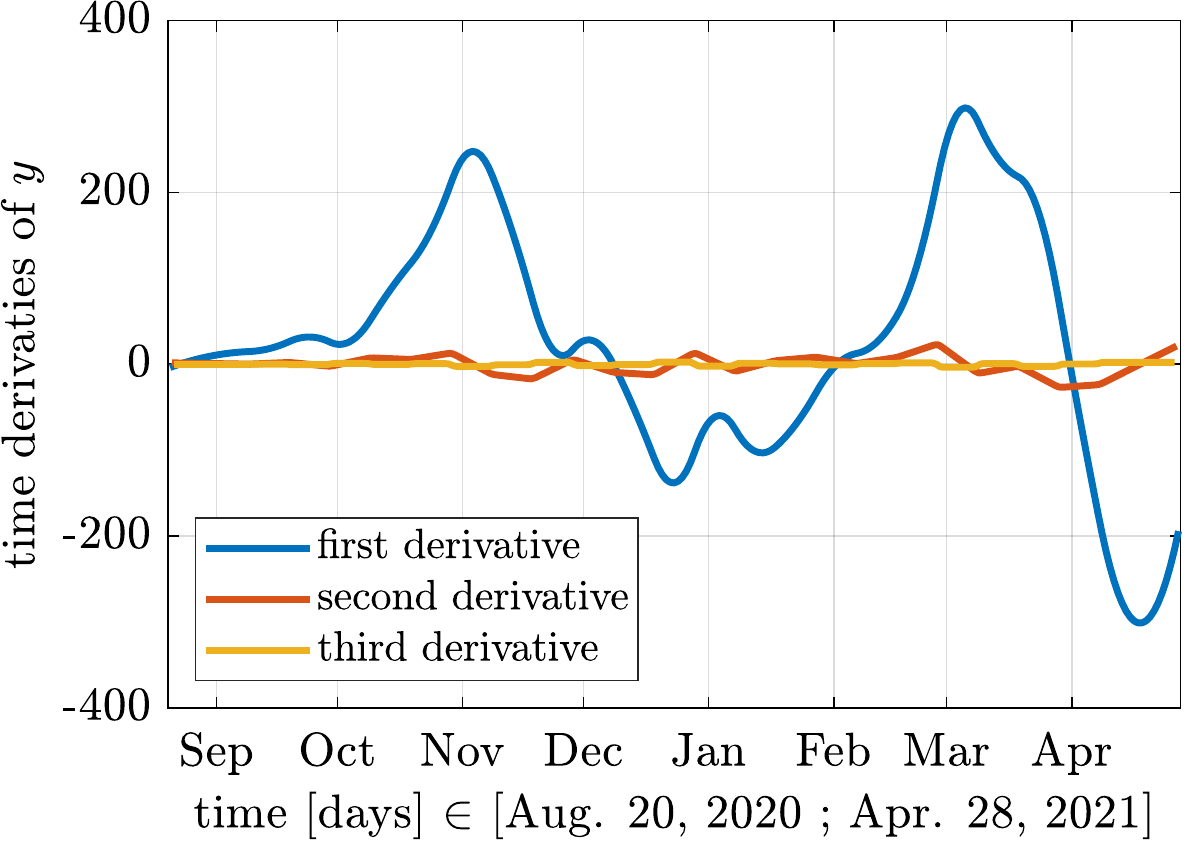}
        \caption{The first three derivatives of the cubic spline interpolated measurement data.}
        \label{fig:derivaties_of_H}
    \end{minipage}
    \hfill
    \begin{minipage}{\tilelength}
        \centering
        \includegraphics[width=\textwidth]{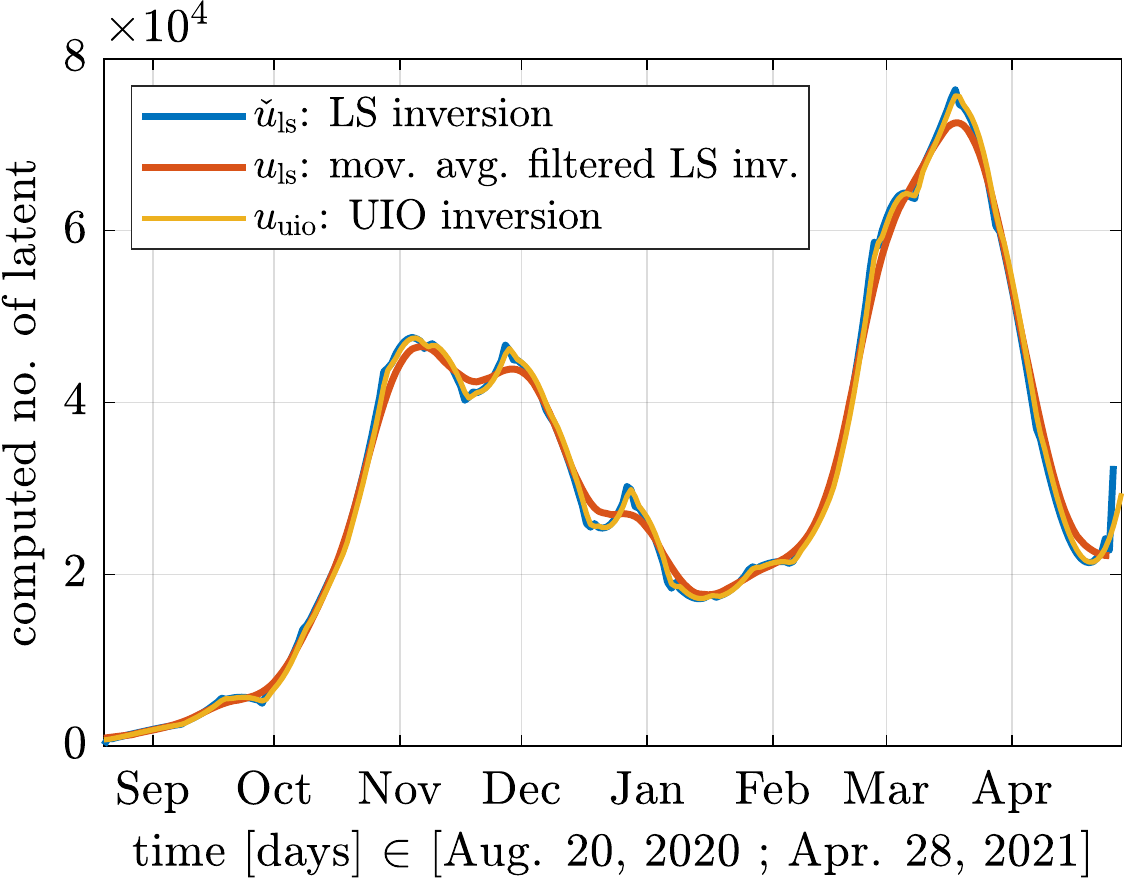}
        \caption{Computed estimate for the number of people in the latent phase using a least squares solution and a dynamical input reconstruction.}
        \label{fig:LS_dyn_inv}
    \end{minipage}
    \hfill
    \begin{minipage}{\tilelength}
        \centering
        \includegraphics[width=\textwidth]{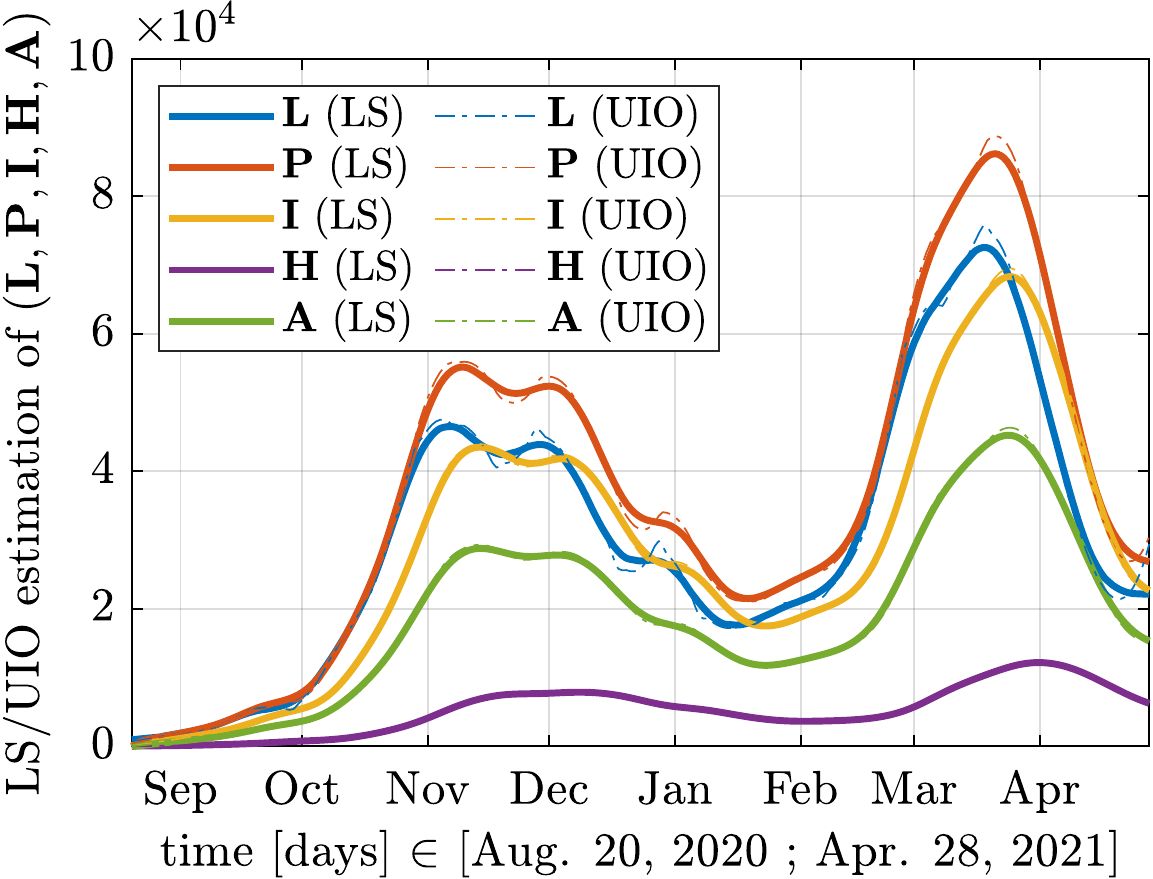}
        \caption{Estimated state variables obtained through the least squares solution $u_{\mathrm{ls}}$ (solid) and by the unknown-input observer (dashed).}
        \label{fig:all_states}
    \end{minipage}
    \hfill
    \begin{minipage}{\tilelength}
        \centering
        \includegraphics[width=\textwidth]{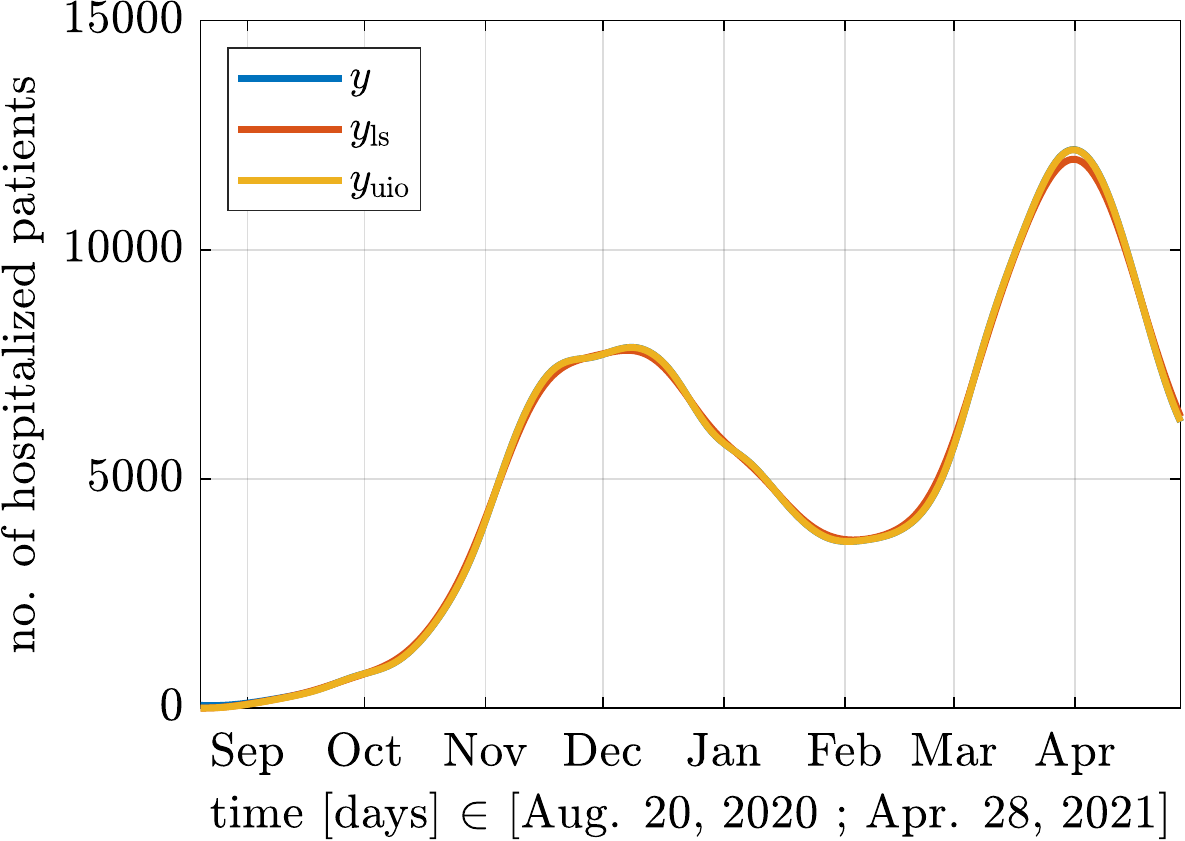}
        \caption{Measured and filtered hospitalization data and its reconstruction using both the least squares solution $u_{\mathrm{ls}}$ and the dynamically computed $u_{\mathrm{uio}}.$}
        \label{fig:hosp_valid}
    \end{minipage}
    \hfill
\begin{minipage}{\tilelength}
        \centering
        \includegraphics[width=\textwidth]{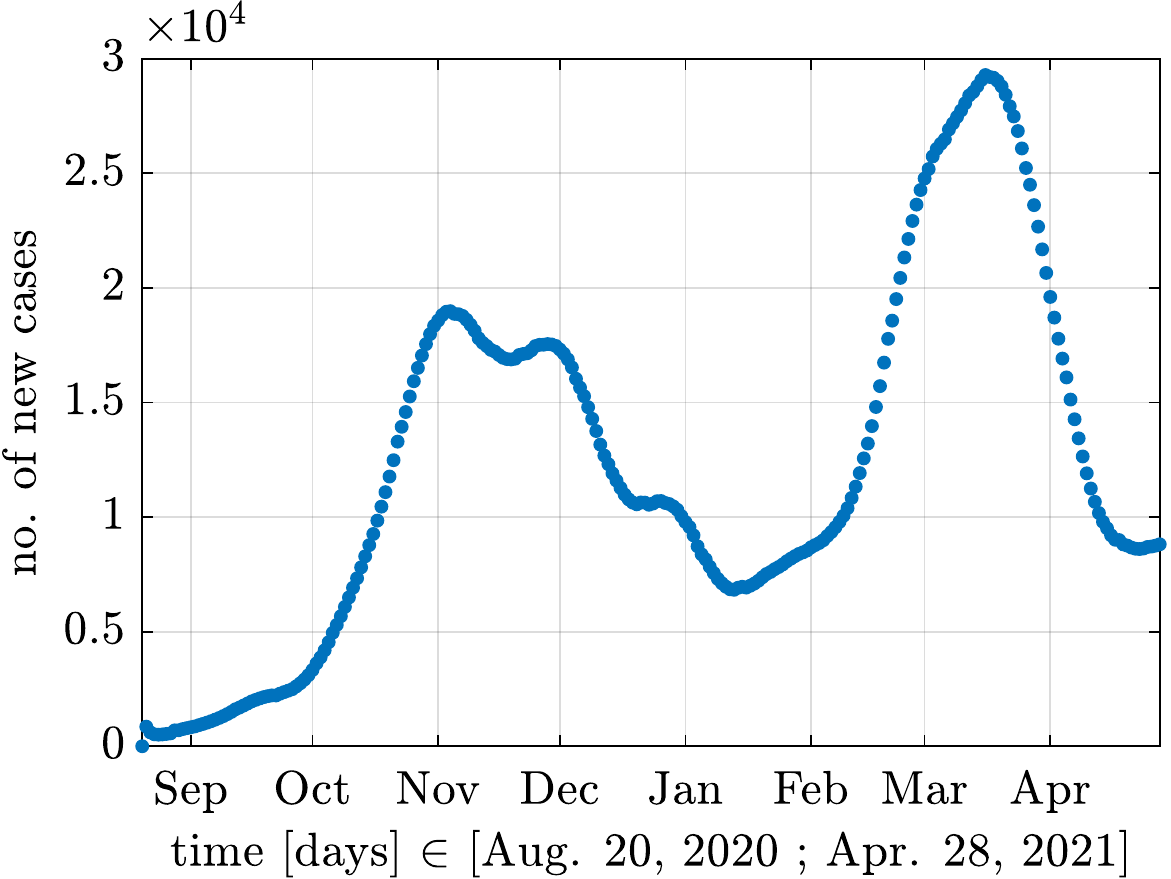}
        \caption{Computed number of new infections per day.}
        \label{fig:daily_new}
    \end{minipage}
\end{figure*}    
\begin{figure*}
    \begin{minipage}{\tilelength}
        \centering
        \includegraphics[width=\textwidth]{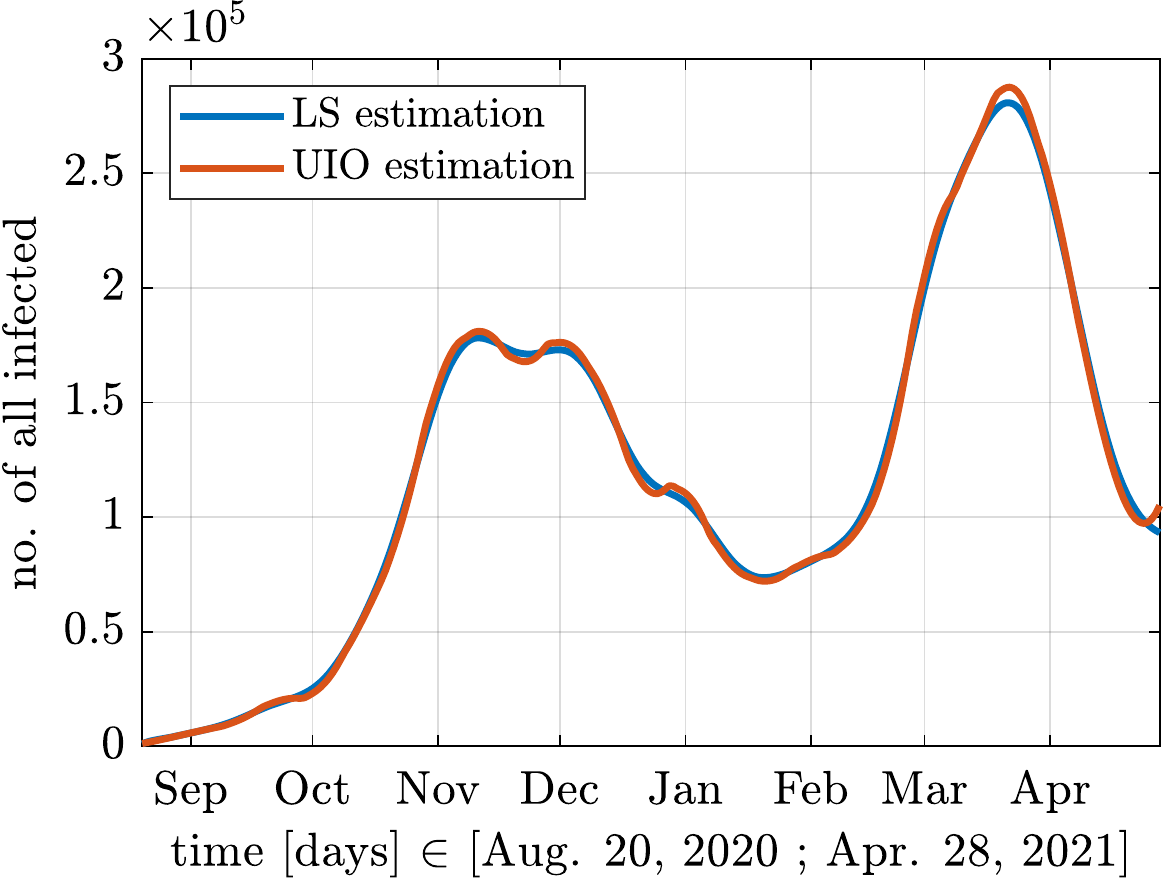}
        \caption{Computed number of all infected people (sum of $\cL$, $\cP$, $\cA$, $\cI$, and $\cH$).}
        \label{fig:all_infected}
    \end{minipage}
    \hfill
\begin{minipage}{\tilelength}
        \centering
        \includegraphics[width=\textwidth]{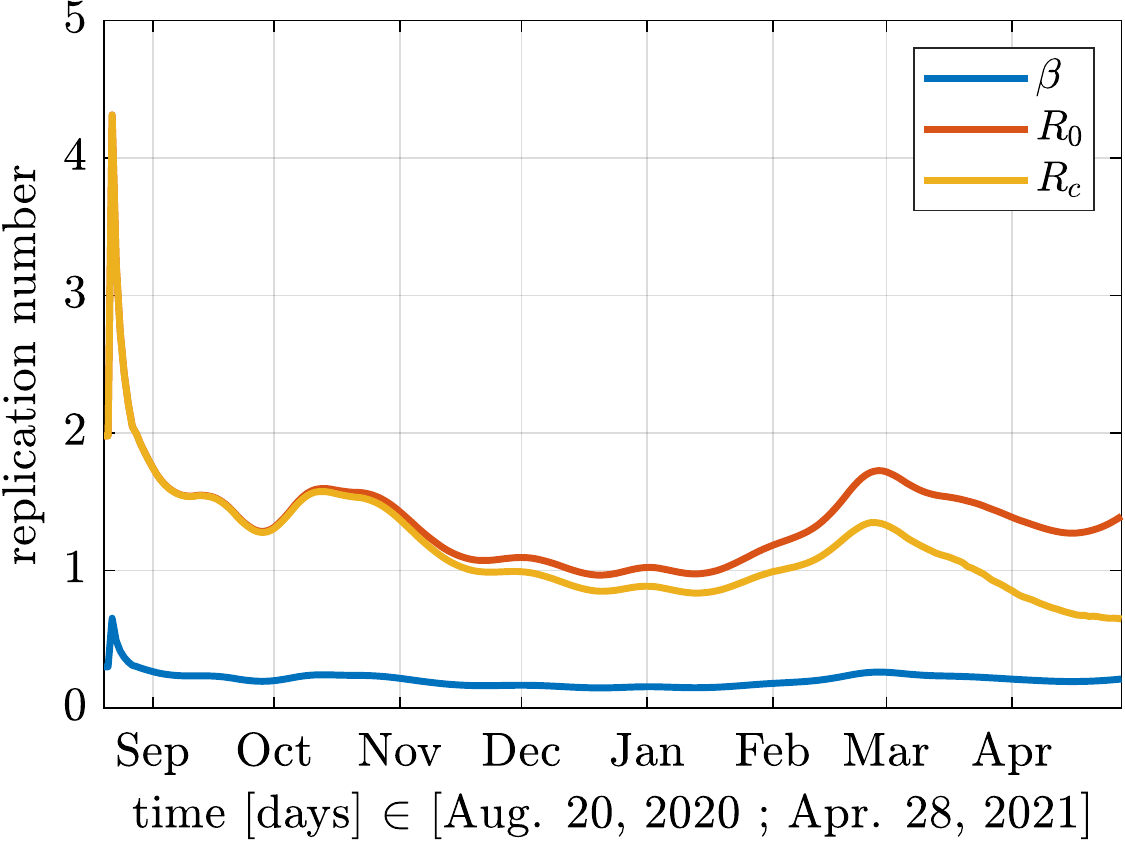}
        \caption{Time evolution of the estimated reproduction number.}
        \label{fig:R0_beta}
    \end{minipage}
\hfill
    \begin{minipage}{\tilelength}
        \centering
        \includegraphics[width=\textwidth]{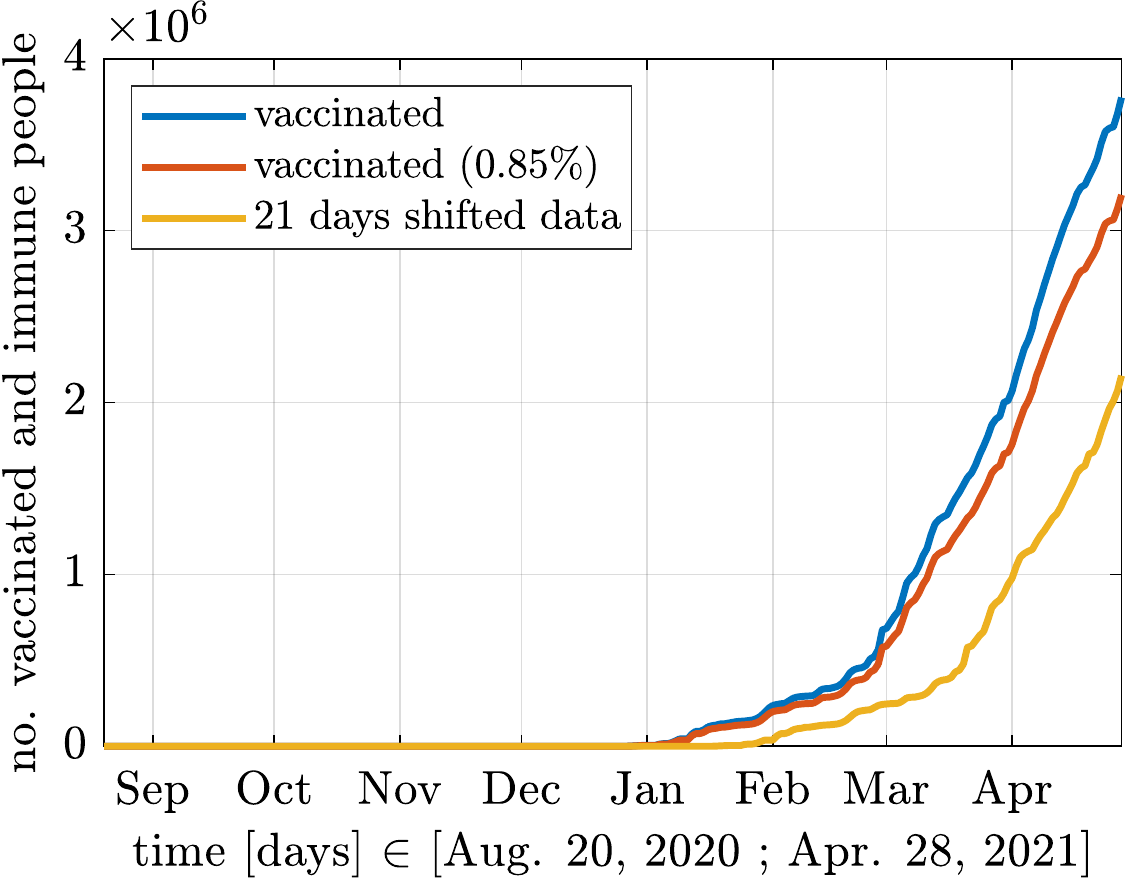}
        \caption{Number of vaccinated and estimated number of truly immune people.}
        \label{fig:vaccinated}
    \end{minipage}
    \hfill
    \begin{minipage}{\tilelength}
        \centering
        \includegraphics[width=\textwidth]{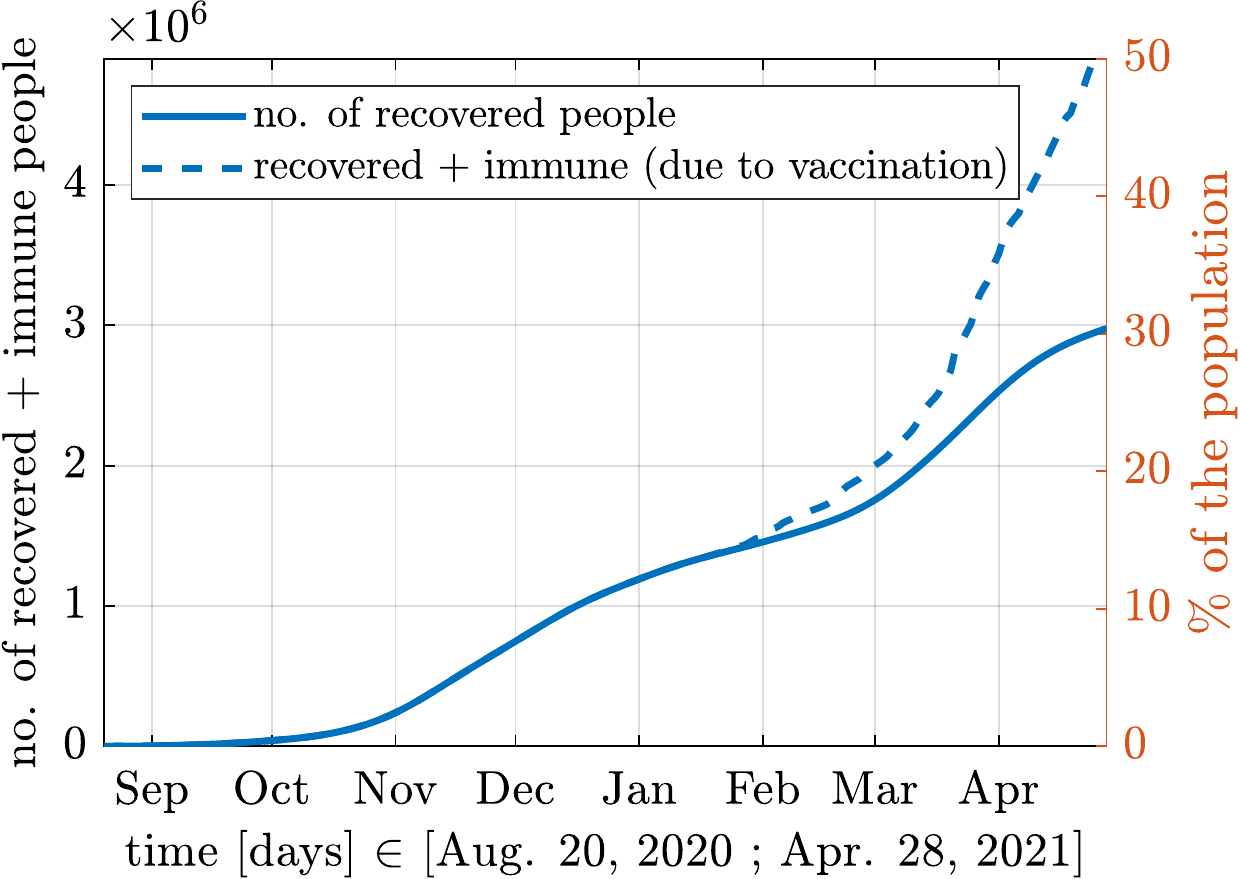}
        \caption{Estimated number of recovered or immune people.}
        \label{fig:recovered_plus_V}
    \end{minipage}
    \hfill
    \begin{minipage}{\tilelength}
        \centering
        \includegraphics[width=\textwidth]{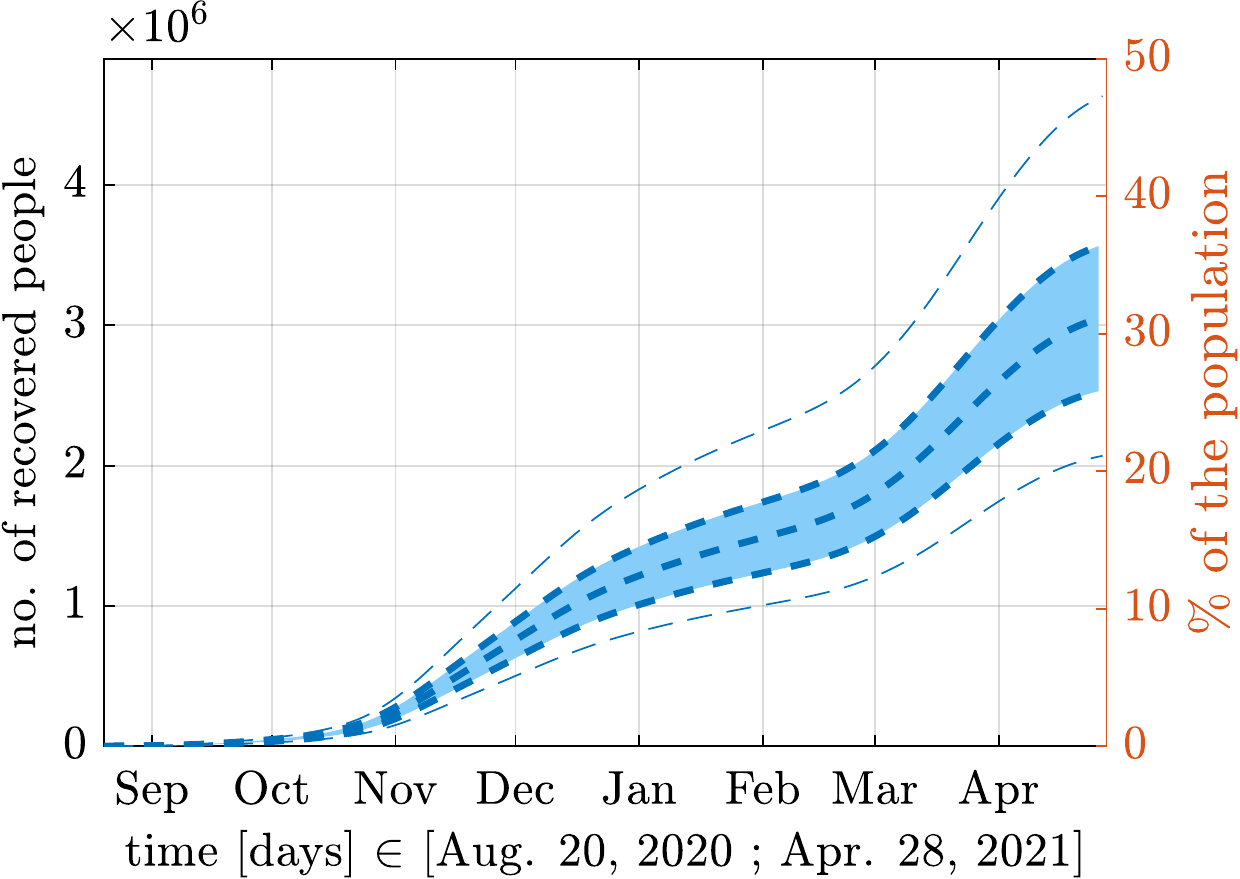}
        \caption{Estimated number of recovered people. The shaded blue area illustrates the standard deviation of the simulated trajectories. The light blue dashed lines show the simulated minimum/maximum number of recovered people.}
        \label{fig:recovered}
    \end{minipage}
    \hfill
\end{figure*}

\end{small}

\color{black}

\subsection{Estimation of the reproduction numbers}
It is visible from Eqs. \eqref{R0} and \eqref{eq:Rct} that the parameter $\beta$ is needed to compute estimates for the reproduction numbers. The right hand side of Eq. \eqref{eq:compS} depends linearly on $\beta$, therefore, a recursive least squares (RLS) algorithm with exponential forgetting can be applied \cite{lennart1999system}. For this, we discretize \eqref{eq:compL} using a simple forward-Euler scheme with sampling time $T_s$ which can be written in linear regression form as
\begin{align}
\underbrace{\cL_{k+1}+ (\alpha T_s -1) \cL_k}_{\pi_k} = \beta_k \underbrace{ T_s (\cP_k + \cI_k + \delta \cA_k)\cS_k / N}_{\varphi_k}.
\end{align}
Then, the RLS algorithm for estimating $\beta$ can be written as
\begin{align}
\hat{\beta}_{k+1} & = \hat{\beta}_k + P_k \varphi_k (\pi_k - \varphi_k  \hat{\beta}_k) \\
P_{k+1} & = (\lambda \cdot P_k^{-1} + \varphi_k^2)^{-1}
\end{align}
where $\hat{\beta}_k$ is the estimate for $\beta$, $P_k$ is the auxiliary variable, and $\lambda=0.9$ is the forgetting factor. The estimates for the reproduction numbers $R_0$ and $R_c$ are given by substituting $\hat{\beta}$ into Eqs. \eqref{R0} and \eqref{eq:Rct}.
We note that the computation of $R_c$ uses the simple vaccination model presented in Remark \ref{TMP_Tovu0ltLdm3r}.

%

\section{Results and Discussion}
The studied period is from August 20, 2020 to April 28, 2021. We considered this time independently of the drastically and successfully suppressed first wave of the epidemic in the spring of 2020.  During the first half of August, the main infection indicators (daily new infections, test positivity, number of hospitalized people) were so low that the zero initial condition assumption for the transfer function representation is acceptable. (We remark that the correct convergence of the state estimates would be ensured even in the case of non-zero initial conditions after a transient period, since the subsystem \eqref{eq:PIH_ss} is asymptotically stable and observable.)

\subsection{Evaluation of the inversion}

In Fig. \ref{fig:LS_dyn_inv}, we illustrate the least squares solution $\check u_{\mathrm{ls}}$ for the unknown input, its { 14-day} long moving average filtered data $u_{\mathrm{ls}}$, and the dynamically reconstructed input $u_{\mathrm{uio}}$.
In Fig. \ref{fig:all_states}, the solid lines illustrate the simulated state variables of the $(\cP,\cI,\cH,\cA)$ subsystem of \eqref{eq:compartmental} driven by the filtered least squares solution $\check u_{\mathrm{ls}}$.
Whereas, the dashed lines of Fig. \ref{fig:all_states}, constitute the observed state variables of $\SigmaPIH$ and \eqref{eq:compA}, and the reconstructed unknown input $u_{\mathrm{uio}}$ in the approximative knowledge of the daily number of hospitalized patients ($y$) and its first three derivatives.

To evaluate the reliability of the state observations, we simulate $\SigmaPIH$ with the computed (and linearly interpolated) unknown input functions $\check u_{\mathrm{ls}}$, $u_{\mathrm{ls}}$, and $u_{\mathrm{uio}}$.
Then, the reconstructed hospitalization data, denoted by $\check y_{\mathrm{ls}}$, $y_{\mathrm{ls}}$, and $y_{\mathrm{uio}}$, respectively, are compared to the actual and the filtered measurements $\check y$, $\bar y$, and $y$.
Fig. \ref{fig:hosp_valid} illustrates that the shape of $y$, $y_{\mathrm{ls}}$, and $y_{\mathrm{uio}}$ are qualitatively the same.

To quantify the difference between the measured and the reconstructed signals,
(instead of the standard squared error model)
we use the following relative squared error formula:
\begin{align}
    \label{eq:Error_evaluation_formula}
d_{\mathrm{r}}(y,\hat{y})
    \!=\!
    \frac{\norm{y \!-\! \hat{y}}_{2}}{\norm{y}_{2}}
    \!=\!
\qty(\sum_{k = 1}^{T-3} \!\abs{ y_k \!-\! \hat{y}_k }^2\!)^{\!\!\frac{1}{2}}
    \!
    \qty(\sum_{k = 1}^{T-3} \!\abs{ y_k }^2\!)^{\!\!\!-\frac{1}{2}}\!\!\!.
\end{align}
Practically, \eqref{eq:Error_evaluation_formula} is the relative distance between the two series obtained from their absolute $\ell_2$ distance $\norm{y - \hat{y}}_2$ and divided by the $\ell_2$ norm $\norm{y}_2$ of the \emph{reference} sequence $y$.
It is worth mentioning that the relative distance between $y$ and the identically zero function is $1$.
Furthermore, note that the last \emph{three} values of the least squares solution $\check u_{\mathrm{ls}}$ cannot be considered, since \eqref{eq:PIH_DT_tf} is a \emph{third}-order DT-LTI model and thus the summation in \eqref{eq:Error_evaluation_formula} is performed only up to $T-3$.

In Table \ref{table:dist_y_y}, we present the (pairwise) relative $\ell_2$ distance between the measurement $\check y$, $\bar y$, $y$
and the reconstructed hospitalization data $\check y_{\mathrm{ls}}$, $y_{\mathrm{ls}}$, $y_{\mathrm{uio}}$.
The results show a slightly smaller relative error when an unknown-input observer is applied.
However, it is worth remarking that the preliminary filtering steps of Section \ref{sec:Comp_preproc} are essential when the derivative-based observer filter \eqref{eq:PIH_uio_ss} is applied.

\begin{table}
\caption{Relative distance of the reconstructed hospitalization data to the actual and filtered data.}
    \label{table:dist_y_y}
    \centering
    \begin{tabular}{|c|ccc|}
        \hline
        $d_{\mathrm{r}}(\downarrow,\rightarrow)$ & $\check y_{\mathrm{ls}}$ & $y_{\mathrm{ls}}$ & $y_{\mathrm{uio}}$
        \\
        \hline
        $\check y$   & 0.02764   &    0.02941    &   0.02541 \\
        $\bar y$     & 0.01476   &    0.01579    &   0.01064 \\
        $y$          & 0.01111   &     0.0128    &   0.00167 \\
\hline
    \end{tabular}
\end{table}

\subsection{Discussion of the obtained results}
Fig. \ref{fig:daily_new} shows the computed number of daily new infections, where two distinct waves (called the 2nd and 3rd waves, respectively) can be observed. The peaks of this curve during November-December 2020 and March 2021 match the available data, and the computation suggests a roughly 4-fold ratio between the model computed and officially detected new cases. The effect of the restrictions (closure of restaurants and gyms, online education in secondary schools, banning of most gatherings, curfew from 8pm to 5am, etc.) introduced in the first half of November is clearly visible. In Fig. \ref{fig:all_infected}, the sum of all the infected compartments ($\cL$, $\cP$, $\cI$, $\cA$, $\cH$) is shown. The obtained numbers for October-November are comparable to the results of a nationwide testing in Slovakia at the end of October 2020, where about 1.06 percent of the population proved to be COVID-positive \cite{holt2021covid}. The estimated reproduction numbers are visible in Fig. \ref{fig:R0_beta}. In February 2021, $R_c$ increased above 1 again which indicates the spread of a more contagious variant (B.1.1.7) causing the large 3rd wave although the November restrictions were not at all relaxed at that time. The ratio of the peaks of the estimated $R_0$ in December 2020 and late February 2021 is 1.618, which is completely in agreement with the 1.4 - 1.8 interval reported in the UK \cite{volz2021transmission}. On March 8, 2021, further measures were introduced including the closing of all schools which clearly contributed to the decrease of the number of daily infections from the second half of March.
Fig. \ref{fig:recovered_plus_V} illustrates the estimated number of recovered people.
The computations suggest that approximately $30$ percent of the Hungarian population might have already gone through the disease.
This result is extended with a straightforward uncertainty analysis (illustrated in Fig. \ref{fig:recovered}).
For each constant model parameter in \eqref{eq:compS} - \eqref{eq:compD} we assumed and uncertainty interval of $\pm20\%$ around the nominal value.
5000 simulation runs were performed with uniformly distributed random parameters taken from the multidimensional uncertainty interval.
The mean value of the sum of the estimated number of recovered people is 4.0 times higher on April 28, 2021 than the official cumulative number of COVID-positive cases mentioned in the Introduction.
This multiplier is comparable to \cite{rocchetti2020estimating}, where the calculated ratio of probable total and detected cases is between 3.93 and 7.94 for 10 European countries not including Hungary, with a mean of 4.85. For this comparison, we have to take into consideration that testing intensity was significantly increased during the 2nd and 3rd waves.

\section{Conclusions}
The data of the COVID-19 pandemic in Hungary was studied in this paper using a systems theoretic approach. The approach is based on the dynamic inversion of a linear subsystem of the used compartmental model. We have shown two possible approaches leading to similar results to track the non-measured state variables, which then allows the estimation of the time dependent reproduction numbers. We remark that the modeling of vaccination is not necessary to give an estimation for the compartments containing infected people, although it is definitely required to estimate the number of susceptible people and the time-dependent reproduction number during the 3rd wave. The described approach not only gives an estimate for the cumulative number of infected and recovered people but it is also able to unravel information on the whole time-course of the epidemic process. The obtained results fall into the ranges that have been published in the literature for different European countries. Future work will be focused on the extension of the method to take model uncertainty into consideration in a systematic way.

\bibliographystyle{IEEEtran}
\bibliography{references}

\end{document}